\documentclass{eptcs}
\providecommand{\event}{FLACOS 2011} 
\usepackage{breakurl}             

\usepackage{latexsym,makeidx,graphicx,color,colortbl}
\usepackage{moreverb,amsmath,amssymb}
\usepackage{pstricks,pst-node}
\usepackage{paralist}

\newcommand{\codiag}{\textit{C-O Diagrams}}

\newtheorem{definition}{Definition}

\newcommand{\bdfn}{\begin{definition} \begin{rm}}
\newcommand{\edfn}{\vspace{-2ex}
{\flushright $\Box$\\
\mbox{}\vspace{-2ex}
} \end{rm} \end{definition}}
\newcommand{\edfnt}{ \end{rm} \end{definition}}
\newcommand{\bthm}{\begin{theorem} \begin{rm}}
\newcommand{\ethm}{\vspace{-4ex}{\flushright $\Box$\\
\mbox{}\vspace{-4ex}} \end{rm} \end{theorem}}
\newcommand{\bprop}{\begin{proposition} \begin{rm}}
\newcommand{\eprop}{\vspace{-4ex}{\flushright $\Box$\\
\mbox{}\vspace{-4ex}} \end{rm} \end{proposition}}
\newcommand{\bcor}{\begin{corollary}\begin{rm}}
\newcommand{\ecor}{\vspace{-4ex}{\flushright $\Box$\\
\mbox{}\vspace{-4ex}} \end{rm} \end{corollary}}
\newcommand{\blem}{\begin{lemma} \begin{rm}}
\newcommand{\elem}{\vspace{-4ex}{\flushright $\Box$\\
\mbox{}\vspace{-4ex}} \end{rm} \end{lemma}}
\newcommand{\bfact}{\begin{fact} \begin{rm}}
\newcommand{\efact}{\vspace{-4ex}{\flushright $\Box$\\
\mbox{}\vspace{-4ex}} \end{rm} \end{fact}}
\newcommand{\bex}{\begin{example} \begin{rm}}
\newcommand{\eex}{\vspace{-2ex}{\flushright $\Box$\\
\mbox{}\vspace{-2ex}} \end{rm} \end{example}}

\newcommand{\nat}{{\rm I\! N}}

\long\def\comment#1{}

\newcommand{\flechald}[2]{\renewcommand{\arraystretch}{0.5}
\begin{array}{c}
{\scriptstyle #1}\\
\longrightarrow\vspace*{-0.1cm}\\
{\scriptstyle #2}
\end{array}
\renewcommand{\arraystretch}{1}
}
\newcommand{\flechalu}[2]{\renewcommand{\arraystretch}{0.5}
\begin{array}{c}
{\scriptstyle #1}\\
\longrightarrow_u\vspace*{-0.1cm}\\
{\scriptstyle #2}
\end{array}
\renewcommand{\arraystretch}{1}
}

\newcommand{\comentario}[1]{}

\title{Timed Automata Semantics for Visual e-Contracts\thanks{Partially supported by the Spanish government (co financed by FEDER founds) with the project TIN2009-14312-C02-02 and the JCCLM regional project PEII09-0232-7745.}}
\author{Enrique Mart\'{i}nez, M. Emilia Cambronero, Gregorio D\'{i}az
\institute{Department of Computer Science\\
University of Castilla-La Mancha\\
Albacete, Spain}
\email{\{emartinez, gregorio, emicp\}@dsi.uclm.es}
\and
Gerardo Schneider
\institute{Department of Computer Science and Engineering\\
Chalmers $\mid$ University of Gothenburg, Sweden\\
Department of Informatics\\
University of Oslo, Norway}
\email{gersch@chalmers.se}
}
\def\titlerunning{Timed Automata Semantics for Visual e-Contracts}
\def\authorrunning{Enrique Mart\'{i}nez, M. Emilia Cambronero, Gregorio D\'{i}az \& Gerardo Schneider}

\begin{document}
\maketitle

\begin{abstract}

C-O Diagrams have been introduced as a means to have a more visual representation of electronic contracts, where it is possible to represent the obligations, permissions and prohibitions of the different signatories, as well as what are the penalties in case of not fulfillment of their obligations and prohibitions. In such diagrams we are also able to represent absolute and relative timing constraints. In this paper we present a formal semantics for C-O Diagrams based on 
timed automata extended with an ordering of states and edges in order to represent different deontic modalities.

\end{abstract}

\section{Introduction}

In the software context, the term {\it contract} has traditionally been used as a metaphor to represent limited kinds of ``agreements'' between software elements at different levels of abstraction. The first use of the term in connection with software programming and design  was done by Meyer in the context of the language Eiffel ({\it programming-by-contracts}, or {\it design-by-contract}) \cite{meyer86dbc}. This notion of contracts basically relies on the Hoare's notion of pre and post-conditions and invariants. Though this paradigm has proved to be useful for developing object oriented systems, it seems to have shortcomings for novel development paradigms such as service-oriented computing and component-based development. These new applications have a more involved interaction and therefore require a more sophisticated notion of contracts.

As a response, behavioural interfaces have been proposed to capture richer properties than simple pre and post-conditions \cite{Hatcliff2009}. Here it is possible to express contracts on the history of events, including causality properties. However, the approach is limited when it comes to contracts containing exceptional behaviour, since the focus is mainly on the interaction concerning expected (and prohibited) behaviour.
  
In the context of SOA, there are different service contract specification languages, like ebXML \cite{ebxml}, WSLA \cite{wsla}, and WS-Agreement~\cite{ws-agreement}. These standards and specification languages suffer from one or more of the following problems: They are restricted to bilateral contracts, lack formal semantics (so it is difficult to reason about them), their treatment of functional behaviour is rather limited and the sub-languages used to specify, for instance, security constraints are usually limited to small application-specific domains. The lack of suitable languages for contracts in the context of SOA is a clear conclusion of the survey \cite{OR08csc} where a taxonomy is presented.

More recently, some researchers have investigated how to adapt deontic logic \cite{McNamara2006} to define (consistent) contracts targeted to software systems where the focus is on the normative notions of obligation, permission and prohibition, including sometimes exceptional cases (e.g., \cite{PS09cl}). 
Independently of the application domain, there still is need to better fill the gap between a contract understood by non-experts in formal methods (for its use), its logical representation (for reasoning), and its internal machine-representation  (for runtime monitoring, and to be manipulated by programmers). We see two possible ways to bridge this gap: i) to develop suitable techniques to get a good translation from contracts written in natural language into formal languages, and ii) to provide a graphical representation (and tools) to manipulate contracts at a high level, with formal semantics supporting automatic translation into the formal language. 
We take in this paper the second approach.

In \cite{MCD+10} we have introduced \codiag, a graphical representation for contracts allowing the representation of complex clauses describing the obligations, permissions, and prohibitions of different signatories (as defined in deontic logic \cite{McNamara2006}), as well as {\it reparations} describing contractual clauses in case of not fulfillment of obligations and prohibitions. Besides, \codiag\ permit to define real-time constraints. In \cite{Martinez2010} some of the satisfaction rules needed to check if a timed automaton satisfies a \textit{C-O Diagram} specification were defined. These rules were originally miscalled ``formal semantics''. The goal of this paper is to further develop our previous work, 
in particular we present here a formal semantics for \codiag\ based on timed automata, extended with an ordering of states and edges.

The rest of the work is structured as follows: Section \ref{Model} presents \codiag\ and their syntax, Section \ref{Semantics} develops the formal semantics of \codiag, including its implementation in UPPAAL \cite{UPPAAL} and a small example. The work is concluded in Section \ref{Conclusions}.

\section{C-O Diagrams Description and Syntax}
\label{Model}

\begin{figure}
\vspace{-0.5cm}
\begin{center}
  \includegraphics[width=4cm]{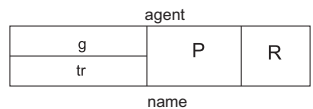}
\end{center}
	\vspace{-0.7cm}
  \caption{Box structure}
  \label{goal}
\vspace{-0.1cm}
\end{figure}

In Fig.~\ref{goal} we show the basic element of \codiag. It is called a \textbf{box} and it is divided into four fields. On the left-hand side of the box we specify the conditions and restrictions. The \textit{guard} \textbf{g} specifies the conditions under which the contract clause must be taken into account (boolean expression). The \textit{time restriction} \textbf{tr} specifies the time frame during which the contract clause must be satisfied (deadlines, timeouts, etc.). The \textit{propositional content} \textbf{P}, on the center, is the main field of the box, and it is used to specify normative aspects (obligations, permissions and prohibitions) that are applied over actions, and/or the specification of the actions themselves. The last field of these boxes, on the right-hand side, is the \textit{reparation} \textbf{R}. This reparation, if specified by the contract clause, is a reference to another contract that must be satisfied in case the main norm is not satisfied (a \textit{prohibition} is violated or an \textit{obligation} is not fulfilled, there is no reparation for \textit{permission}), considering the clause eventually satisfied if this reparation is satisfied. Each box has also a name and an agent. The \textit{name} is useful both to describe the clause and to reference the box from other clauses, so it must be unique. The \textit{agent} indicates who is the performer of the action.

\begin{figure*}
\vspace{-0.1cm}
\begin{center}
  \includegraphics[width=14cm]{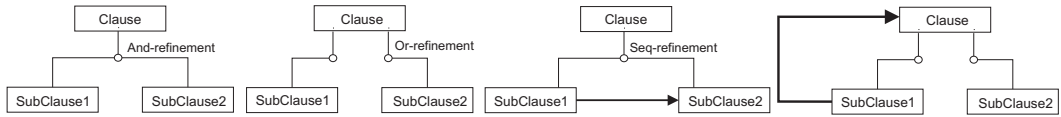}
\end{center}
	\vspace{-0.7cm}
  \caption{AND/OR/SEQ refinements and repetition in \codiag}
  \label{refinem}
\vspace{-0.5cm}
\end{figure*}

These basic elements of \codiag\ can be refined by using AND/OR/SEQ refinements, as shown in Fig.~\ref{refinem}. The aim of these refinements is to capture the hierarchical clause structure followed by most contracts. An \textbf{AND-refinement} means that all the subclauses must be satisfied in order to satisfied the parent clause. An \textbf{OR-refinement} means that it is only necessary to satisfy one of the subclauses in order to satisfy the parent clause, so as soon as one of its subclauses is fulfilled, we conclude that the parent clause is fulfilled as well. A \textbf{SEQ-refinement} means that the norm specified in the target box (\textit{SubClause2} in Fig.~\ref{refinem}) must be fulfilled after satisfying the norm specified in the source box (\textit{SubClause1} in Fig.~\ref{refinem}). By using these structures we can build a hierarchical tree with the clauses defined by a contract, where the leaf clauses correspond to the atomic clauses, that is, to the clauses that cannot be divided into subclauses. There is another structure that can be used to model \textbf{repetition}. This structure is represented as an arrow going from a subclause to one of its ancestor clauses (or to itself), meaning the repetitive application of all the subclauses of the target clause after satisfying the source subclause. For example, in the right-hand side of Fig.~\ref{refinem}, we have an \textbf{OR-refinement} with an arrow going from \textit{SubClause1} to \textit{Clause}. It means that after satisfying \textit{SubClause1} we apply \textit{Clause} again, but not after satisfying \textit{SubClause2}.

\begin{figure}
\vspace{-0.5cm}
\begin{center}
  \includegraphics[width=9cm]{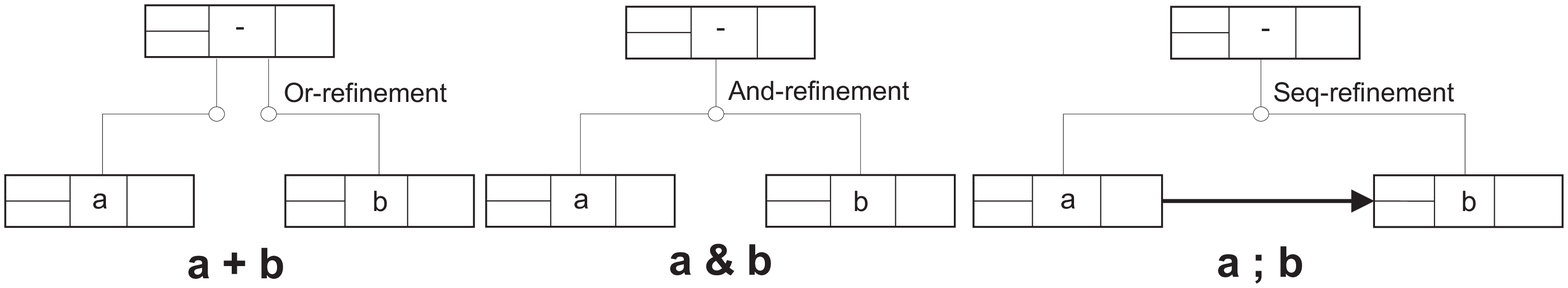}
\end{center}
	\vspace{-0.7cm}
  \caption{Compound actions in \textit{C-O Diagrams}}
  \label{CompoundActions}
\vspace{-0.1cm}
\end{figure}

It is only considered the specification of \textit{atomic actions} in the \textbf{P} field of the leaf boxes of our diagrams. The composition of actions can be achieved by means of the different kinds of refinement. In this way, an AND-refinement can be used to model \textit{concurrency} ``\&'' between actions, an OR-refinement can be used to model a \textit{choice} ``+'' between actions, and a SEQ-refinement can be used to model \textit{sequence} ``;'' of actions. In Fig.~\ref{CompoundActions} we can see an example about how to model these compound actions through refinements, given two atomic actions \textit{a} and \textit{b}.

The \textit{deontic norms} (obligations, permissions and prohibitions) that are applied over these actions can be specified in any box of our \textit{C-O Diagrams}, affecting all the actions in the leaf boxes that are descendants of this box. If it is the case that the box where we specify the deontic norm is a leaf, the norm only affects the atomic action we have in this box. It is used an upper case ``\textit{O}'' to denote an obligation, an upper case ``\textit{P}'' to denote a permission, and an upper case ``\textit{F}'' to denote a prohibition (forbidden). These letters are written in the top left corner of field \textbf{P}.

\begin{figure}
\vspace{-0.1cm}
\begin{center}
  \includegraphics[width=9cm]{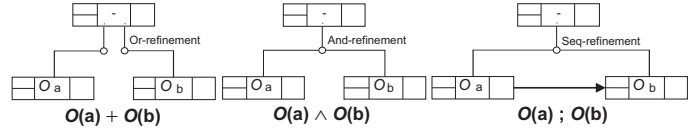}
\end{center}
	\vspace{-0.7cm}
  \caption{Composition of norms in \textit{C-O Diagrams}}
  \label{CompoundNorms}
\vspace{-0.5cm}
\end{figure}

The composition of deontic norms is also achieved by means of the different refinements we have in \textit{C-O Diagrams}. Thus, an AND-refinement corresponds to the \textit{conjunction} operator ``$\wedge$'' between norms, an OR-refinement corresponds to the \textit{choice} operator ``$+$'' between norms, and a SEQ-refinement corresponds to the \textit{sequence} operator ``$;$'' between norms. For example, we can imagine having a leaf box specifying the obligation of performing an action \textit{a}, written as \textit{O}(a), and another leaf box specifying the obligation of performing an action \textit{b}, written as \textit{O}(b). These two norms can be combined in the three different ways mentioned before through the different kinds of refinement (Fig.~\ref{CompoundNorms}). However, the specification of deontic norms in our diagrams must fulfill the following rule: exactly one deontic norm must be specified in each one of the branches of our hierarchical tree, i.e., we cannot have an action without a deontic norm applied over it and we cannot have deontic norms applied over other deontic norms. We have also that \textit{agents} are only specified in the boxes where a deontic norm is defined, being each agent associated to a concrete deontic norm. Finally, the \textit{repetition} of both, actions and deontic norms, can be achieved by means of the repetition structure we have in \codiag.

We have given here an abridged description of \codiag. A more detail description can be found in \cite{MCD+10}, including a qualitative and quantitative evaluation, and a discussion on related work.

\begin{definition}\label{def1} (\codiag\ Syntax) 
We consider a finite set of real-valued variables $\mathcal{C}$ standing for clocks, a finite set of non-negative integer-valued variables $\mathcal{V}$, a finite alphabet $\Sigma$ for atomic actions, a finite set of identifiers $\mathcal{A}$ for agents, and another finite set of identifiers $\mathcal{N}$ for names. The greek letter $\epsilon$ means that and expression is not given, i.e., it is empty.

We use $C$ to denote the contract modelled by a \textit{C-O Diagram}. The diagram is defined by the following EBNF grammar:

{

\begin{center}
\begin{tabular}{rcl}
    $C$ & $:=$ &  \begin{array}[t]{l}
						      (agent,name,g,tr,O(C_2),R) \,|									
						      \end{array} \\
    &  &  \begin{array}[t]{l}		
		      (agent,name,g,tr,P(C_2),\epsilon) \,|	
		      \end{array} \\
		&  &  \begin{array}[t]{l}		
		      (agent,name,g,tr,F(C_2),R) \,|
		      \end{array} \\
		&  &  \begin{array}[t]{l}		
					(\epsilon,name,g,tr,C_1,\epsilon)
		      \end{array} \\		
    $C_1$ & $:=$ &  \begin{array}[t]{l}
								      C \,(And \; C)^+\,|\,
								      C \,(Or \; C)^+\,|\,
								      C \,(Seq \; C)^+
								      \end{array} \\
		$C_2$ & $:=$ &  \begin{array}[t]{l}
											a \,|\,
								      C_3 \,(And \; C_3)^+\,|\,
								      C_3 \,(Or \; C_3)^+\,|\,
								      C_3 \,(Seq \; C_3)^+
								      \end{array} \\
    $C_3$ & $:=$ &  \begin{array}[t]{l}
								      (\epsilon,name,\epsilon,\epsilon,C_2,\epsilon)
								      \end{array} \\
		$R$ & $:=$ &  \begin{array}[t]{l}
						      C \,|\,
						      \epsilon
						      \end{array} \\		
\end{tabular}
\end{center}
}

\noindent where $a \in \Sigma$, $agent \in \mathcal{A}$ and $name \in \mathcal{N}$. Guard $g$ is $\epsilon$ or a conjunctive formula of atomic constraints of the form: $v \sim n$ or $v - w \sim n$, for $v,w\in \mathcal{V}$, $\sim \in \{\, \leq, <, =, >, \geq \, \}$ and $n \in \nat$, whereas timed restriction $tr$ is $\epsilon$ or a conjunctive formula of atomic constraints of the form: $x \sim n$ or $x -y \sim n$, for $x,y \in \mathcal{C}$, $\sim \in \{\, \leq, <, =, >, \geq \, \}$ and $n \in \nat$. $O$, $P$ and $F$ are the deontic operators corresponding to obligation, permission and prohibition, respectively, where $O(C_2)$ states the obligation of performing $C_2$, $F(C_2)$ states prohibition of performing $C_2$, and $P(C_2)$ states the permission of performing $C_2$. $And$, $Or$ and $Seq$ are the operators corresponding to the refinements we have in \codiag, AND-refinement, OR-refinement and SEQ-refinement, respectively.

\end{definition}

The simplest contract we can have in \textit{C-O Diagrams} is that composed of only one box including the elements $agent$ and $name$. Optionally, we can specify a guard $g$ and a time restriction $tr$. We also have a deontic operator ($O$, $P$ or $F$) applied over an atomic action $a$, and in the case of obligations and prohibitions it is possible to specify another contract $C$ as a reparation.

We use $C_1$ to define a more complex contract where we combine different deontic norms by means of any of the different refinements we have in \textit{C-O Diagrams}. In the box where we have the refinement into $C_1$ we cannot specify an agent nor a reparation because these elements are always related to a single deontic norm, but we still can specify a guard $g$ and a time restriction $tr$ that affect all the deontic norms we combine.

Once we write a deontic operator in a box of our diagram, we have two possibilities as we can see in the specification of $C_2$: we can just write a simple action $a$ in the box, being the deontic operator applied only over it, or we can refine this box in order to apply the deontic operator over a compound action. In this case we have that the subboxes ($C_3$) cannot define a new deontic operator as it has already been defined in the parent box (affecting all the subboxes).

\section{\codiag\ Semantics}
\label{Semantics}

The \codiag\ semantics is defined by means of a transformation into a {\it Network of Timed Automata}\, (NTA), that is defined as a set of timed automata \cite{Alur,Alur94} that run simultaneously, using the same set of clocks and variables, and synchronizing on the common actions.

In what follows we consider a finite set of real-valued variables
$\mathcal{C}$ ranged over by $x,y,\ldots$ standing for clocks,
a finite set of non-negative integer-valued variables $\mathcal{V}$, ranged
over by $v,w,\ldots$\,
and a finite alphabet $\Sigma$ ranged over by $a,b,\ldots$
standing for actions. We will use letters $r,r',\ldots$ to denote
sets of clocks. We will denote by ${\it Assigns}$ the set of possible
assignments,
${\it Assigns} = \{ v:={\it expr} \,|\, v \in \mathcal{V}\}$,
where {\it expr} are arithmetic expressions using naturals
and variables.
Letters $s,s' \ldots$ will be used to represent a set
of assignments.

A {\it guard or invariant condition}\, is
a conjunctive formula of atomic constraints of the form:
$x \sim n$, $x -y \sim n$,
$v \sim n$ or $v - w \sim n$,
for $x,y \in \mathcal{C}$, $v,w\in \mathcal{V}$,
$\sim \in \{\, \leq, <, =, >, \geq \, \}$ and $n \in \nat$.
The set of guard or invariant conditions will be denoted by $\mathcal{G}$,
ranged over by $g, g', \ldots$.

\begin{definition}\label{defTA} (Timed Automaton) \\
A {\em timed automaton} is a tuple $(N, n_0, E, I)$, where
$N$ is a finite set of locations (nodes),
$n_0 \in N$ is the initial location,
$E \subseteq N \times \mathcal{G} \times \Sigma \times
\mathcal{P}({\it Assigns}) \times
2^\mathcal{C} \times N$ is the set of edges, where
the subset of {\em urgent edges} is called $E_u \subseteq E$, and
they will graphically be distinguished as they will have
their arrowhead painted in white.
$I \,: \, N \rightarrow \mathcal{G}$ is a function
that assigns invariant conditions (which could be empty)
to locations.

\end{definition}

From now on, we will write
$n \flechald{g,a,r}{s} n'$ to denote
$(n,g,a,s,r,n')\in E$, and
$n \flechalu{g,a,r}{s} n'$ when
\linebreak
$(n,g,a,s,r,n')\in E_u$.

In an NTA we distinguish two types of actions: internal and synchronization actions. Internal actions can be executed by the corresponding automata independently, and they will be ranged over the letters $a,b\ldots$. Synchronization actions, however, must be executed simultaneously by two automata, and they will be ranged over letters $m, m',\ldots$ and come from the synchronization of two actions $m!$ and $m?$, executed from two different automata. Due to the lack of space, we refer the reader to \cite{UPPAAL2} for the definition of the semantics of timed automaton and NTA.

To specify the \codiag\ semantics, we add the definition of two orderings, $\prec_N$ and $\prec_E$, where:

\begin{compactitem}

\item $\prec_N$ is a (strict, partial) ordering on N where $n \prec_N n'$ means that node $n$ is {\em better} than node $n'$.

\item $\prec_E$ is a (strict, partial) ordering on E where $e \prec_N e'$ means that edge $e$ is {\em better} than edge $e'$.

\end{compactitem}

We also add a \textbf{violation set} $V(n)$ associated to each node $n$ in $N$, that is the set of contractual obligations and prohibitions that are violated in $n$.

\begin{definition}\label{def2} (Violation Set) 
Let us consider the set of contractual obligations and prohibitions $CN$ ranged over $cn$, $cn'$,\ldots\ standing for identifiers of obligations and prohibitions. We write $n \not \models cn$ to express that obligation or prohibition $cn$ is violated in node $n$. Therefore, the {\em violation set} is defined as $V(n)=\{cn \, | \, cn \in CN$ and  $n \not \models cn\}$.
\end{definition}

Another set called \textbf{satisfaction set} $S(n)$ is also associated to each node $n$ in $N$. This set is composed by the contractual obligations and prohibitions that have already been satisfied in $n$. 

\begin{definition}\label{def3} (Satisfaction Set) 
Let us consider the set of contractual obligations and prohibitions $COF$ ranged over $cof$, $cof'$,\ldots\ standing for identifiers of obligations and prohibitions. We write $n \models cof$ to express that obligation or prohibition $cof$ has been satisfied in node $n$ (we consider a prohibition satisfied in node $n$ if it has not been violated and cannot be violated anymore because the time frame specified for the prohibition has expired). Hence, the {\em satisfaction set} is defined as $S(n)=\{cof \, | \, cof \in COF$ and  $n \models cof\}$.
\end{definition}

Once these two sets have been defined, we can formally define the \textbf{ordering on nodes} $\prec_N$, by comparing the violation sets and the satisfaction sets of the nodes, and the \textbf{ordering on edges} $\prec_E$, by comparing the violation sets and the satisfaction sets of the target nodes of the edges.

\begin{definition}\label{def4} (Ordering on Nodes) 
A {\em node $n_1$ is better than another node $n_2$} if the violation set of $n_1$ is a proper subset of the violation set of $n_2$ or, if the violation sets are the same, a node $n_1$ is better than another node $n_2$ if the satisfaction set of $n_1$ is a proper superset of the satisfaction set of $n_2$, that is, $n_1 \prec_N n_2$ iff $(V(n_1) \subset V(n_2))$ or $(V(n_1) = V(n_2)$ and $S(n_1) \supset S(n_2))$.
\end{definition}

\begin{definition}\label{def5} (Ordering on Edges) 
An {\em edge $e_1$ is better than another edge $e_2$} if the source node is the same in both cases but the violation set of the target node of $e_1$ is a proper subset of the violation set of the target node of $e_2$ or, if the violation sets are the same, an edge $e_1$ is better than another edge $e_2$ if the satisfaction set of the target node of $e_1$ is a proper superset of the satisfaction set of the target node of $e_2$. Considering $e_1=(n_1,g_1,a_1,s_1,r_1,n{_1}')$ and $e_2=(n_2,g_2,a_2,s_2,r_2,n{_2}')$, $e_1 \prec_E e_2$ iff $(n_1=n_2)$ and $(V(n{_1}') \subset V(n{_2}')$ or $(V(n{_1}') = V(n{_2}')$ and $S(n{_1}') \supset S(n{_2}')))$.
\end{definition}

Finally, another set called \textbf{permission set} $P(n)$ is associated to each node $n$ in $N$. This set influences neither the ordering on nodes nor the ordering on edges, it is used just to record the permissions in the contract that have been made effective.

\begin{definition}\label{def6} (Permission Set) 
Let us consider the set of contractual permissions $CP$ ranged over $cp$, $cp'$,\ldots\ standing for identifiers of permissions. We write $n \models cp$ to express that permission $cp$ has already been made effective in node $n$. Then, the {\em permission set} is defined as $P(n)=\{cp \, | \, cp \in CP$ and  $n \models cp\}$.
\end{definition}

Graphically, when we draw a timed automaton extended with these three sets, we write under each node $n$ between braces its violation set $V(n)$ on the left, its satisfaction set $S(n)$ on the centre and its permission set $P(n)$ on the right. In the initial node of the automata we build corresponding to \codiag\ these three sets are empty. By default, a node keeps in these sets the same content of the previous node when we compose the automata. Only in a few cases the content of these sets is modified (when an obligation or a prohibition is violated, an obligation or a prohibition is satisfied or a permission is made effective).

Concerning the \textbf{real-time restrictions} $tr$ specified in the contract, the two types of time restrictions we can have in \codiag\ must be translated in a different way for their inclusion into a timed automaton construction:

\begin{compactitem}

\item A time restriction specified using \textbf{absolute time} must be specified in timed automata by rewriting the terms in which absolute time references occur. For that purpose we define a global clock $T \in \mathcal{C}$ that is never reset during the execution of the automata and, taking into account the moment at which the contract is enacted, we rewrite the absolute time references as deadlines involving clock $T$ and considering the smallest time unit needed in the contract. For example, let us consider a clause that must be satisfied between the \textit{5th of November} and the \textit{10th of November}, and that the contract containing this clause is enacted the \textit{31st of October}. If we suppose that \textit{days} is the smallest time unit used in the contract for the specification of real-time restrictions, the time restriction of this clause is written as $(T \geq 5) \, and \, (T \leq 10)$.

\item A time restriction specified using \textbf{relative time} must be specified in timed automata by introducing an additional clock to register the amount of time that has elapsed since another clause has been satisfied, resetting the additional clock value when this happens and specifying the deadline using it. We call this clock $t_{name}$, where $name$ is the clause used as reference for the specification of the time restriction. Therefore, we define a set of additional clocks $C_{add}=\{t_{name} \, | \,  t_{name} \in \mathcal{C}\}$ including a clock for every clause that is used as reference in the time restriction of at least another clause. For example, let us consider a contract with a clause that must be satisfied between \textit{5} and \textit{10} days after another clause $name1$ has been satisfied. In this case we define an additional clock $t_{name1}$ that is reset to zero when clause $name1$ is satisfied ($t_{name1}:=0$) and the time restriction of the other clause is written as $(t_{name1} \geq 5) \, and \, (t_{name1} \leq 10)$.

\end{compactitem}

As a result, the set of clocks of the timed automata would be $\mathcal{C} = \{T\} \cup C_{add}$. When we construct the timed automata corresponding to \codiag, we always consider $(x \geq t1) \, and \, (x \leq t2)$ as the interval corresponding to the time restriction $tr$ of the clause, where $x \in \mathcal{C}$ is the clock used for its specification ($x=T$ in the case of absolute time and $x=t_{name}$ in the case of relative time, being $name$ the clause used as reference), $t1 \in \nat$ is the beginning of the interval and $t2 \in \nat$ is the end of the interval ($t1 \leq t2$). If $tr$ does not define the lower bound of the interval we take $t1=0$, if $tr$ does not define the upper bound of the interval we take $t2=\infty$, and if $tr = \epsilon$ we take $t1=0$, $t2=\infty$ and $x=T$.

Once we have given these extensions of the definition of timed automata and we have explained how the different kinds of time restriction can be expressed, considering all the different elements we can specify in a \textit{C-O Diagram}, we can define the transformation of the diagrams into timed automata by induction using several transformation rules.

\begin{definition}\label{def7} (\codiag\ Transformation Rules: Part I)
{\renewcommand{\labelenumi}{(\arabic{enumi})}
\begin{enumerate}

\item An \textbf{atomic action} in a \textit{C-O Diagram}, that is, $(\epsilon,name,\epsilon,\epsilon,a,\epsilon)$ corresponds to the timed automaton ${A}=(N_{{A}}, n_{0_{{A}}}, E_{{A}}, I_{{A}})$, where:

\begin{compactitem}

\item $N_{{A}}=\{a_{init},a_{end}\}$.

\item $n_{0_{{A}}}=a_{init}$.

\item $E_{{A}}=\{a_{init} \flechald{a}{} a_{end}\}$.

\item $I_{{A}}=\emptyset$.

\end{compactitem}

The violation (\textbf{V}), satisfaction (\textbf{S}) and permission (\textbf{P}) sets are not modified, so $V(a_{init})=V(a_{end})$, $S(a_{init})=S(a_{end})$ and $P(a_{init})=P(a_{end})$. This timed automaton can be seen in Fig.~\ref{automaton_A1-2-3-4} \textbf{(A)}.

\begin{figure}
\vspace{-0.5cm}
\begin{center}
  \includegraphics[width=16cm]{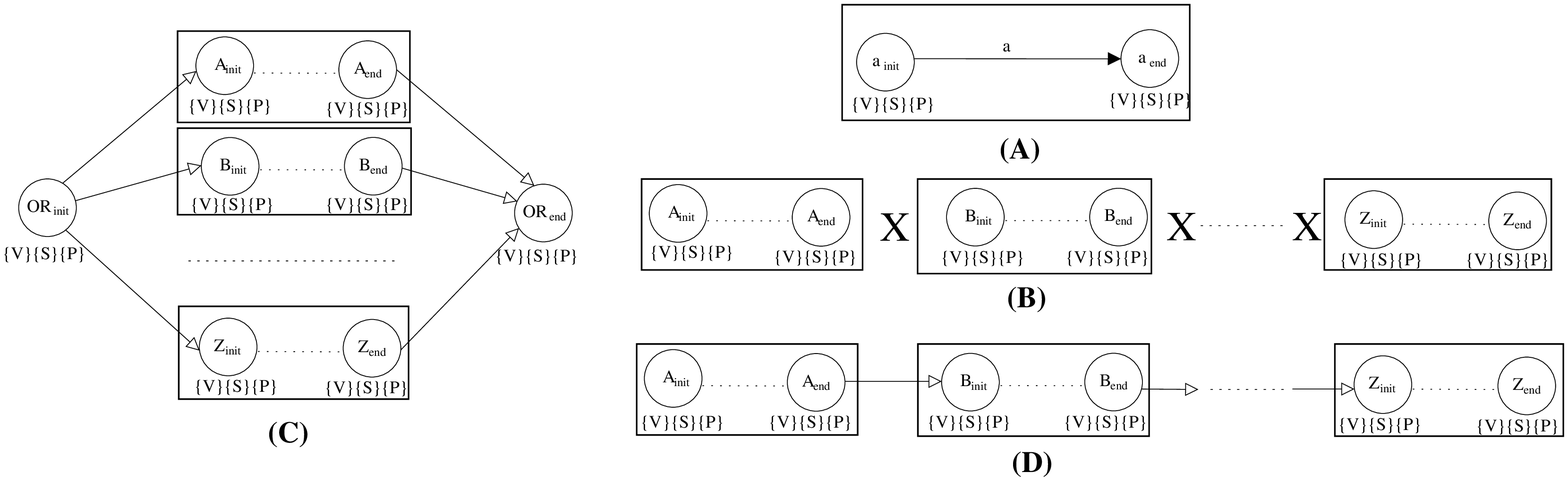}
\end{center}
	\vspace{-0.5cm}
  \caption{Automata corresponding to a \textbf{simple action} $a$  and to \textbf{compound actions}}
  \label{automaton_A1-2-3-4}
\vspace{-0.5cm}
\end{figure}

\item A \textbf{compound action} in a \textit{C-O Diagram} where an \textbf{AND-refinement} is used to compose actions, that is, $(\epsilon,name,\epsilon,\epsilon,C_1 \, And \, C_2 \, And \, \ldots \, And \, C_n,\epsilon)$ corresponds to the cartesian product of the automata corresponding to each one of the subcontracts. Let us consider ${A}, {B}, \ldots, {Z}$ the automata corresponding to the subcontracts $C_1, C_2, \ldots, C_n$ (the actions specified in these subcontracts can be atomic actions or other compound actions). The resulting automaton $AND$ corresponds to the cartesian product of these automata, that is, $AND = {A} \times {B} \times \ldots \times {Z}$. Again, the violation (\textbf{V}), satisfaction (\textbf{S}) and permission (\textbf{P}) sets are not modified, so they are the same in all the nodes. This composition of timed automata is shown graphically in Fig.~\ref{automaton_A1-2-3-4} \textbf{(B)}.

\item A \textbf{compound action} in a \textit{C-O Diagram} where an \textbf{OR-refinement} is used to compose actions, that is, $(\epsilon,name,\epsilon,\epsilon,C_1 \, Or \, C_2 \, Or \, \ldots \, Or \, C_n,\epsilon)$ corresponds to a new automaton in which the automata corresponding to each one of the subcontracts is considered as an alternative. Let us consider ${A}, {B}, \ldots, {Z}$ the automata corresponding to the subcontracts $C_1, C_2, \ldots, C_n$ (the actions specified in these subcontracts can be atomic actions or other compound actions). The resulting automaton $OR$ preserves the structure of the automata we are composing but adding a new initial node $OR_{init}$ and connecting this node by means of urgent edges performing no action to the initial nodes of ${A}, {B}, \ldots, {Z}$ ($A_{init}, B_{init}, \ldots, Z_{init}$). It is also added a new ending node $OR_{end}$ and urgent edges performing no action from the ending nodes of ${A}, {B}, \ldots, {Z}$ ($A_{end}, B_{end}, \ldots, Z_{end}$) to this new ending node. Let ${A}=(N_{{A}}, n_{0_{{A}}}, E_{{A}}, I_{{A}}), {B}=(N_{\mathcal{B}}, n_{0_{{B}}}, E_{{B}}, I_{{B}}), \ldots, {Z}=(N_{{Z}}, n_{0_{{Z}}}, E_{{Z}}, I_{{Z}})$. The resulting automaton is therefore $OR=(N_{OR}, n_{0_{OR}}, E_{OR}, I_{OR})$, where:

\begin{compactitem}

\item $N_{OR}=N_{{A}} \cup N_{{B}} \cup \ldots \cup N_{{Z}}\cup \{OR_{init},OR_{end}\}$.

\item $n_{0_{OR}}=OR_{init}$.

\item $E_{OR}=E_{{A}} \cup E_{{B}} \cup \ldots \cup E_{{Z}} \cup \{OR_{init} \flechalu{}{} A_{init}, OR_{init} \flechalu{}{} B_{init}, \ldots, OR_{init} \flechalu{}{} Z_{init}\} \cup$\\
$\{A_{end} \flechalu{}{} OR_{end}, B_{end} \flechalu{}{} OR_{end}, \ldots, Z_{end} \flechalu{}{} OR_{end}\}$.

\item $I_{OR}=I_{{A}} \cup I_{{B}} \cup \ldots \cup I_{{Z}}$.

\end{compactitem}

The violation (\textbf{V}), satisfaction (\textbf{S}) and permission (\textbf{P}) sets are not modified, so they are the same in all the nodes. This composition of timed automata is shown graphically in Fig.~\ref{automaton_A1-2-3-4} \textbf{(C)}.

\item A \textbf{compound action} in a \textit{C-O Diagram} where a \textbf{SEQ-refinement} is used to compose actions, that is, $(\epsilon,name,\epsilon,\epsilon,C_1 \, Seq \, C_2 \, Seq \ldots Seq \, C_n,\epsilon)$ corresponds to a new automaton in which the automata corresponding to each one of the subcontracts are connected in sequence. Let us consider ${A}, {B}, \ldots, {Z}$ the automata corresponding to the subcontracts $C_1, C_2, \ldots, C_n$ (the actions specified in these subcontracts can be atomic actions or other compound actions). The resulting automaton $SEQ$ preserves the structure of the automata we are composing, adding no extra nodes. We only connect with an urgent edge performing no action the ending node of each automaton in the sequence ($A_{end}, B_{end}, \ldots, Y_{end}$) with the initial node of the next automaton in the sequence ($B_{init}, C_{init}, \ldots, Z_{init}$). This rule is not applied in the cases of $A_{init}$ (as there is not previous ending node to connect) and $Z_{end}$ (as there is not following initial node to connect). Let ${A}=(N_{{A}}, n_{0_{{A}}}, E_{{A}}, I_{{A}}), \mathcal{B}=(N_{{B}}, n_{0_{{B}}}, E_{{B}}, I_{{B}}), \ldots, {Z}=(N_{{Z}}, n_{0_{{Z}}}, E_{{Z}}, I_{{Z}})$. The resulting automaton is therefore $SEQ=(N_{SEQ}, n_{0_{SEQ}}, E_{SEQ}, I_{SEQ})$, where:

\begin{compactitem}

\item $N_{SEQ}=N_{{A}} \cup N_{{B}} \cup \ldots \cup N_{{Z}}$.

\item $n_{0_{SEQ}}=A_{init}$.

\item $E_{SEQ}=E_{{A}} \cup E_{{B}} \cup \ldots \cup E_{{Z}} \cup \{A_{end} \flechalu{}{} B_{init}, B_{end} \flechalu{}{} C_{init}, \ldots, Y_{end} \flechalu{}{} Z_{init}\}$.

\item $I_{SEQ}=I_{{A}} \cup I_{{B}} \cup \ldots \cup I_{{Z}}$.

\end{compactitem}

Again, the violation (\textbf{V}), satisfaction (\textbf{S}) and permission (\textbf{P}) sets are not modified, so they are the same in all the nodes. This composition of timed automata is shown graphically in Fig.~\ref{automaton_A1-2-3-4} \textbf{(D)}.

\end{enumerate}
}
\end{definition}

Until now, we have seen how the automata corresponding to the different actions (atomic or compound) specified in a \textit{C-O Diagram} are constructed and we have seen that these translations do not modify the content of any of the sets (violation, satisfaction or permission). Next, we define the transformation rules specifying how these ``action'' automata are modified when we apply a deontic norm (obligation, permission or prohibition) over the actions in the \textit{C-O Diagram}.

\setcounter{definition}{6}
\begin{definition} (\codiag\ Transformation Rules: Part II)
{\renewcommand{\labelenumi}{(\arabic{enumi})}
\begin{enumerate}

\setcounter{enumi}{4}

\item The application of an \textbf{obligation}, a \textbf{permission} or a \textbf{prohibition} over an action in a \textit{C-O Diagram}, i.e., $(agent,name,g,tr,O/P/F(C),R)$ corresponds to an automaton where the obligation/prohibition of performing the action specified in the subcontract $C$ can be skipped, fulfilled or violated, whereas the permission of performing the action can be skipped, made effective or not made effective. Let us consider ${A}=(N_{{A}}, n_{0_{{A}}}, E_{{A}}, I_{{A}})$ the automaton corresponding to $C$, being $A_{init}$ the initial node and $A_{end}$ the ending node. The resulting automaton ${D(A)}$, where $D \in \{O,P,F\}$, preserves the structure of the automaton ${A}$ but adding a new ending node $A_{time}$ including the obligation over the action in its violation set, the prohibition over the action in its satisfaction set or nothing if a permission over the action is considered. If guard condition $g \neq \epsilon$, we add another ending node $A_{skip}$ where the violation, satisfaction and permission sets are not modified. We also include the obligation over the action in the satisfaction set of $A_{end}$, the prohibition over the action in the violation set of $A_{end}$, or the permission over the action in the permission set of $A_{end}$. An invariant $x \leq t2 + 1$ is added to each node of ${A}$ except $A_{end}$ and each edge performing one of the actions in this automaton is guarded with $(x \geq t1) \, and \, (x \leq t2)$ and action performed by $agent$. New edges guarded with $x = t2 + 1$ and no action performed are added from each node of ${A}$ except $A_{end}$ to the new node $A_{time}$ and, if guard condition $g \neq \epsilon$, an urgent edge from $A_{init}$ to $A_{skip}$ is also added guarded with the guard condition of the clause negated ($\neg g$). Finally, if $t_{name} \in \mathcal{C}$, all the edges reaching $A_{end}$ reset $t_{name}$ in the cases of obligation and permission, whereas all the edges reaching $A_{time}$ reset $t_{name}$ in the case of prohibition. Considering the more complex case, where $g \neq \epsilon$ and $t_{name} \in \mathcal{C}$, and having that $g_1 \equiv (x \geq t1) \, and \, (x \leq t2)$ and $g_2 \equiv x = t2 + 1$, the resulting automaton is therefore ${D(A)}=(N_{{D(A)}}, n_{0_{{D(A)}}}, E_{{D(A)}}, I_{{D(A)}})$, where:

\begin{compactitem}

\item $N_{{D(A)}}=N_{{A}} \cup \{A_{time},A_{skip}\}$.

\item $n_{0_{{D(A)}}}=A_{init}$.

\item $E_{{D(A)}}= \{A_{init} \flechalu{\neg g}{} A_{skip}\} \cup
	\left\{
	  \begin{array}{ll}
		 n \xrightarrow[]{g_1,agent(a)} n' \, | \, n \flechald{a}{} n' \in E_{{A}}$ and $n' \neq A_{end},\\
		 n \xrightarrow[]{g_1,agent(a),t_{name}} n' \, | \, n \flechald{a}{} n' \in E_{{A}}$ and $n'=A_{end},\\
		 n \flechald{g_2}{} A_{time} \, | \, n \in N_{{A}}-\{A_{end}\} & \mathrm{if\ } D=O \\
		 n \xrightarrow[]{g_1,agent(a)} n' \, | \, n \flechald{a}{} n' \in E_{{A}}$ and $n' \neq A_{end},\\
		 n \xrightarrow[]{g_1,agent(a),t_{name}} n' \, | \, n \flechald{a}{} n' \in E_{{A}}$ and $n'=A_{end},\\
		 n \flechald{g_2}{} A_{time} \, | \, n \in N_{{A}}-\{A_{end}\} & \mathrm{if\ } D=P\\
		 n \xrightarrow[]{g_1,agent(a)} n' \, | \, n \flechald{a}{} n' \in E_{{A}},\\
		 n \xrightarrow[]{g_2,t_{name}} A_{time} \, | \, n \in N_{{A}}-\{A_{end}\}& \mathrm{if\ } D=F \\
	  \end{array}
	\right.
$

\item $I_{{D(A)}}=I_{{A}} \cup \{I(n) \equiv x \leq t2 + 1 \, | \, n \in N_{{A}}-\{A_{end}\}\}$.

\end{compactitem}

The resulting timed automata are shown graphically in Fig.~\ref{automaton_A5-6-7-R}, where \textbf{(A)} corresponds to obligation, \textbf{(B)} corresponds to permission and \textbf{(C)} corresponds to prohibition. We consider $a$ one of the atomic actions included in the subcontract $C$.

\begin{figure}
\vspace{-0.5cm}
\begin{center}
  \includegraphics[width=15cm]{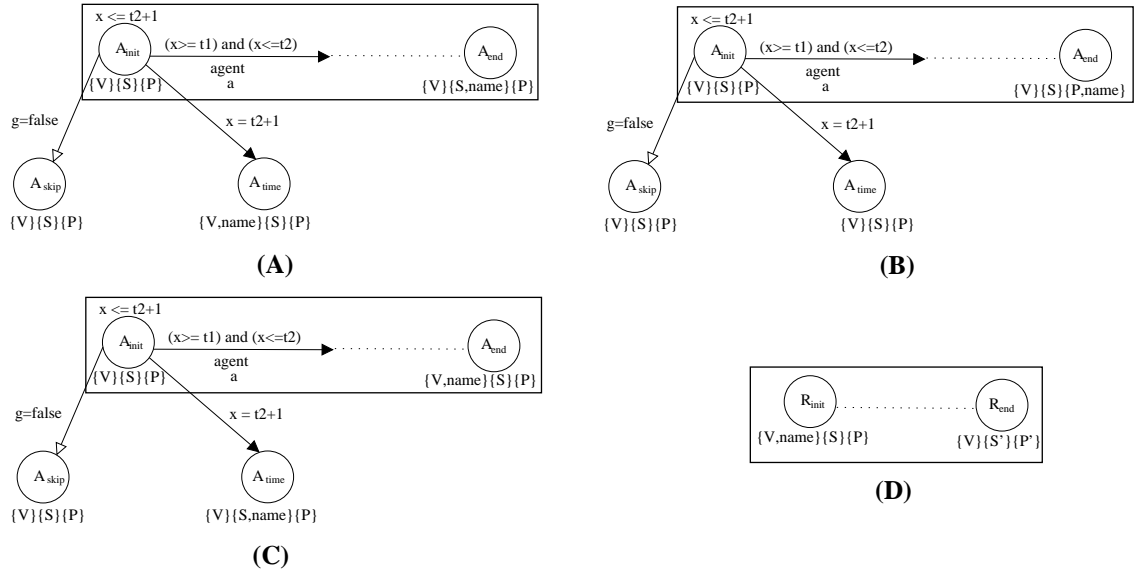}
\end{center}
	\vspace{-0.5cm}
  \caption{Automata corresponding to \textbf{deontic norms} and automaton corresponding to a \textbf{reparation}}
  \label{automaton_A5-6-7-R}
\vspace{-0.5cm}
\end{figure}

\end{enumerate}

}
\end{definition}

We can see that the above constructions can include a reparation contract $R$ in the cases of obligation and prohibition. If this reparation is defined, we have to construct the automaton corresponding to the reparation contract and integrate this automaton as part of the automaton we have generated for the obligation or prohibition. This reparation contract removes the obliged or prohibited clause $name$ from the violation set of the corresponding automaton, as we can see in Fig.~\ref{automaton_A5-6-7-R} \textbf{(D)}.

\setcounter{definition}{6}
\begin{definition} (\codiag\ Transformation Rules: Part III)
{\renewcommand{\labelenumi}{(\arabic{enumi})}
\begin{enumerate}

\setcounter{enumi}{5}

\item An \textbf{obligation} or \textbf{prohibition} in a \textit{C-O Diagram} specifying a contract \textbf{reparation} $R \neq \epsilon$ corresponds to the obligation automaton ${O(A)}$ or the prohibition automaton ${F(A)}$ together with the reparation automaton ${R}$, considering the node with $name$ in its violation ($A_{vio}$) set as the initial node of the reparation automaton ($R_{init}$). In the ending node of the reparation automaton ($R_{end}$) $name$ is removed from the violation set, as the violation has been repaired. In this node we also have that the satisfaction set and the permission set are different from the ones we have in the initial node of the reparation because we have to include in the satisfaction set all the obligations and prohibitions satisfied in the reparation contract, and in the permission set all the permissions that have been made effective in the reparation contract. Let us consider ${D(A)}=(N_{{D(A)}}, n_{0_{{D(A)}}}, E_{{D(A)}}, I_{{D(A)}})$, where $D \in \{O,F\}$, and ${R}=(N_{{R}}, n_{0_{{R}}}, E_{{R}}, I_{{R}})$. The resulting automaton is therefore ${D(A)_R}=(N_{{D(A)_R}}, n_{0_{{D(A)_R}}}, E_{{D(A)_R}}, I_{{D(A)_R}})$, where:

\begin{compactitem}

\item $N_{{D(A)_R}}=N_{{D(A)}} \cup N_{{R}}-\{R_{init}\}$.

\item $n_{0_{{D(A)_R}}}=A_{init}$.

\item $E_{{D(A)_R}}=E_{{D(A)}} \cup \{n \flechald{g,a,r}{s} n' \, | \, n \flechald{g,a,r}{s} n' \in E_{{R}}$ and $n \neq R_{init}\} \cup$\\
$\{A_{vio} \flechald{g,a,r}{s} n' \, | \, n \flechald{g,a,r}{s} n' \in E_{{R}}$ and $n=R_{init}\}$.

\item $I_{{D(A)_R}}=I_{{D(A)}}-\{I(A_{vio})\} \cup \{I(n)  \, | \, n \in  N_{{R}}-\{R_{init}\}\} \cup \{I(A_{vio}) \equiv I(R_{init})\}$.

\end{compactitem}

\end{enumerate}

}
\end{definition}

Finally, we have to define the rules about how the automata corresponding to different deontic norms are composed when we have a composition of deontic norms in our \textit{C-O Diagram}. To make this composition possible, first we need to have only one ending node in the automata corresponding to the different deontic norms. Therefore, we add a new ending node in these automata and urgent edges from the old ending nodes to this new node. Notice that in the case of obligation and prohibition, if there is no reparation defined, the node violating the norm is a final node of the whole automaton construction where the contract is breached. In the case of permission, as no reparation is defined, we have that $P(A)_R=P(A)$.

\setcounter{definition}{6}
\begin{definition} (\codiag\ Transformation Rules: Part IV)
{\renewcommand{\labelenumi}{(\arabic{enumi})}
\begin{enumerate}

\setcounter{enumi}{6}

\item Let ${D(A)_R}=(N_{{D(A)_R}}, n_{0_{{D(A)_R}}}, E_{{D(A)_R}}, I_{{D(A)_R}})$, where $D \in \{O,P,F\}$, be the automaton corresponding to an \textbf{obligation}, a \textbf{prohibition} or a \textbf{permission} in a \textit{C-O Diagram}, specifying a \textbf{reparation} $R \neq \epsilon$ in the two first cases. The corresponding automaton with only one ending node, that we call $A_{final}$ and preserves the violation, satisfaction and permission sets of the previous node, is ${D(A)'_R}=(N_{{D(A)'_R}}, n_{0_{{D(A)'_R}}}, E_{{D(A)'_R}}, I_{{D(A)'_R}})$, where:

\begin{compactitem}

\item $N_{{D(A)'_R}}=N_{{D(A)_R}} \cup \{A_{final}\}$.

\item $n_{0_{{D(A)'_R}}}=n_{0_{{D(A)_R}}}$.

\item $E_{{D(A)'_R}}=E_{{D(A)_R}} \cup \{A_{skip} \flechalu{}{} A_{final}\} \cup 
	\left\{
	  \begin{array}{ll}
		 A_{end} \flechalu{}{} A_{final},R_{end} \flechalu{}{} A_{final} & \mathrm{if\ } D=O \\
		 A_{end} \flechalu{}{} A_{final},A_{time} \flechalu{}{} A_{final} & \mathrm{if\ } D=P\\
		 A_{time} \flechalu{}{} A_{final},R_{end} \flechalu{}{} A_{final} & \mathrm{if\ } D=F \\
	  \end{array}
	\right.
$

\item $I_{{D(A)'_R}}=I_{{D(A)_R}}$.

\end{compactitem}

\end{enumerate}

}
\end{definition}

Therefore, the composition of the automata corresponding to different deontic norms is defined by three additional transformation rules.

\setcounter{definition}{6}
\begin{definition} (\codiag\ Transformation Rules: Part V)
{\renewcommand{\labelenumi}{(\arabic{enumi})}
\begin{enumerate}

\setcounter{enumi}{7}

\item If several norms are composed by an \textbf{AND-refinement}, that is, we have specified the diagram $(\epsilon,name,g,tr,C_1 \, And \, C_2 \, And \ldots And \, C_n,\epsilon)$, their composition corresponds to a network of automata in which we consider all the norms we are composing in parallel. Let us consider $\mathcal{C}_1, \mathcal{C}_2, \ldots, \mathcal{C}_n$ the automata corresponding to the norms we are composing. The resulting network of automata preserves the structure of the automata we are composing, adding to each one of them the additional nodes and edges necessary for synchronization (these nodes are called $C_{init}$ and $C_{final}$ in the first automaton, $C_{isyn}$ and $C_{isyn'}, i=1,\ldots,n-1$ in the other automata). Before its initial node, each automaton synchronizes with the other automata and it synchronizes again after its final node by means of urgent channels ($m_1, m_2, \ldots, m_{n-1}$). In the first automaton we add another node $C_{skip}$ if guard condition of the parent clause $g \neq \epsilon$ and an urgent edge from $C_{init}$ to this new node guarded with the guard condition negated ($\neg g$). In the final node of the first automaton the violation, satisfaction and permission sets are the union of the sets resulting in each one of the automata running in parallel, so we have that Vfinal = ${V1 \cup V2 \cup \ldots \cup Vn}$, Sfinal = ${S1 \cup S2 \cup \ldots \cup Sn}$ and Pfinal = ${P1 \cup P2 \cup \ldots \cup Pn}$. If time restriction of the parent clause $tr \neq \epsilon$, we consider this additional time restriction in all the composed automata together with their own time restrictions. Let $\mathcal{C}_1=(N_{\mathcal{C}_1}, n_{0_{\mathcal{C}_1}}, E_{\mathcal{C}_1}, I_{\mathcal{C}_1}), \mathcal{C}_2=(N_{\mathcal{C}_2}, n_{0_{\mathcal{C}_2}}, E_{\mathcal{C}_2}, I_{\mathcal{C}_2}), \ldots,
\mathcal{C}_n=(N_{\mathcal{C}_n}, n_{0_{\mathcal{C}_n}}, E_{\mathcal{C}_n}, I_{\mathcal{C}_n})$. Considering the case where $g \neq \epsilon$ and $tr \neq \epsilon$, and having that $E_{\mathcal{C}_1}*, E_{\mathcal{C}_2}*, \ldots ,E_{\mathcal{C}_n}*$ are the sets of edges considering time restriction $tr$ together with their own time restriction, the resulting network of automata is therefore $\mathcal{C*}_i=(N_{\mathcal{C*}_i}, n_{0_{\mathcal{C*}_i}}, E_{\mathcal{C*}_i}, I_{\mathcal{C*}_i})$, $i=1,\ldots,n$ where:

\begin{compactitem}

\item $N_{\mathcal{C*}_i}=N_{\mathcal{C}_i} \cup 
	\left\{
	  \begin{array}{ll}
		 C_{init},C_{final},C_{skip} & \mathrm{if\ } i=1 \\
		 C_{isyn},C_{isyn'},C_{i-1syn},C_{i-1syn'} & \mathrm{if\ } i=2,\ldots,n-1 \\
		 C_{i-1syn},C_{i-1syn'} & \mathrm{if\ } i=n
	  \end{array}
	\right.
$

\item $n_{0_{\mathcal{C*}_i}}=
	\left\{
	  \begin{array}{ll}
		 C_{init} & \mathrm{if\ } i=1 \\
		 C_{i-1syn},C_{i-1syn'} & \mathrm{if\ } i=2,\ldots,n \\		 
	  \end{array}
	\right.
$

\item $E_{\mathcal{C*}_i}=E_{\mathcal{C}_i} \cup 
	\left\{
	  \begin{array}{ll}
		 C_{init} \flechalu{\neg g}{} C_{skip}, C_{init} \flechald{m_{i}!}{} C_{iinit},\\
		 C_{ifinal} \flechald{m_{i}!}{} C_{final} & \mathrm{if\ } i=1 \\
		 C_{i-1syn} \flechald{m_{i-1}?}{} C_{isyn}, C_{isyn'} \flechald{m_{i-1}?}{} C_{i-1syn'},\\
		 C_{isyn} \flechald{m_{i}!}{} C_{iinit}, C_{ifinal} \flechald{m_{i}!}{} C_{isyn'} & \mathrm{if\ } i=2,\ldots,n-1 \\
		 C_{i-1syn} \flechald{m_{i-1}?}{} C_{iinit}, C_{ifinal} \flechald{m_{i-1}?}{} C_{final} & \mathrm{if\ } i=n
	  \end{array}
	\right.
$

\item $I_{\mathcal{C*}_i}=I_{\mathcal{C}_i} \cup \{I(n) \equiv x \leq t2 + 1 \, | \, n \in N_{\mathcal{C}_i}-\{C_{ifinal}\}\}$.

\end{compactitem}

This composition of timed automata is shown graphically in Fig.~\ref{automaton_A8-9-10} \textbf{(A)}.

\begin{figure}
\vspace{-0.5cm}
\begin{center}
  \includegraphics[width=16cm]{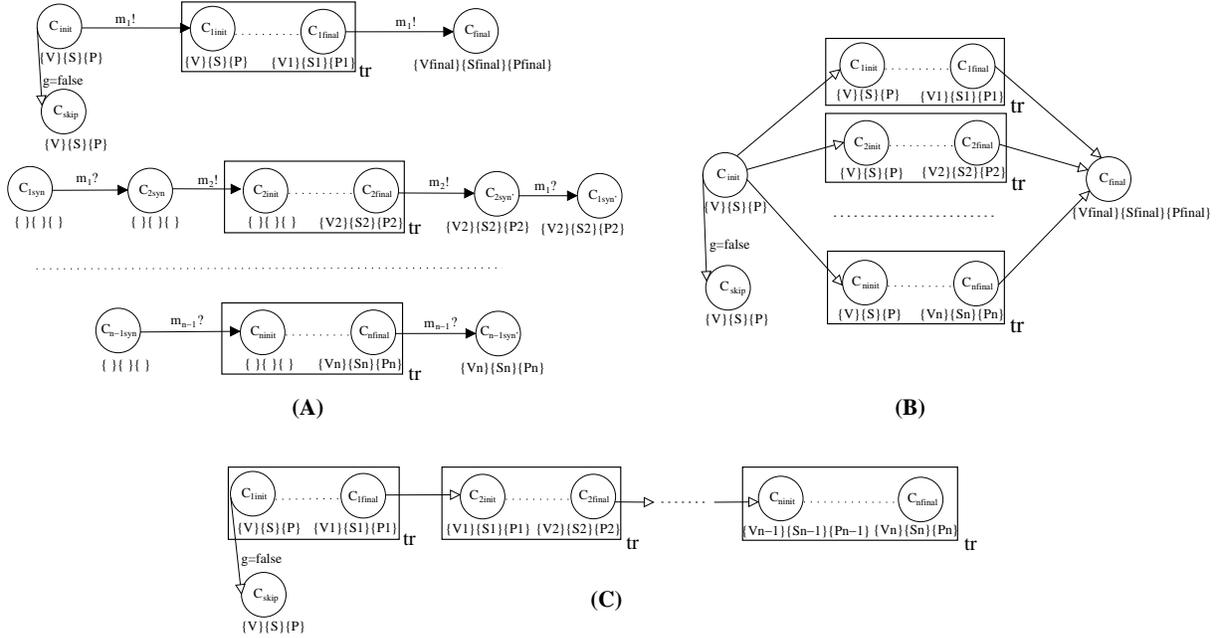}
\end{center}
	\vspace{-0.5cm}
  \caption{Automata corresponding to the \textbf{compositions of deontic norms}}
  \label{automaton_A8-9-10}
\vspace{-0.5cm}
\end{figure}

\item If several norms are composed by an \textbf{OR-refinement}, that is, we have specified the diagram $(\epsilon,name,g,tr,C_1 \, Or \, C_2 \, Or \ldots Or \, C_n,\epsilon)$, their composition corresponds to an automaton in which the automata corresponding to each one of the norms is considered as an alternative. Let us consider $\mathcal{C}_1, \mathcal{C}_2, \ldots, \mathcal{C}_n$ the automata corresponding to the norms we are composing. The resulting automaton $OR*$ preserves the structure of the automata we are composing, adding two nodes called $C_{init}$ and $C_{final}$. We define an urgent edge performing no action for each one of the norms we are composing connecting $C_{init}$ with the initial node of the automaton corresponding to the norm and we also define an urgent edge performing no action for each one of the norm we are composing connecting the final node of its automaton with $C_{final}$. We add another node $C_{skip}$ if guard condition of the parent clause $g \neq \epsilon$ and an urgent edge from $C_{init}$ to this new node guarded with the guard condition negated ($\neg g$). In the final node of this new structure we keep the violation, satisfaction and permission sets of the previous final node, so we have that Vfinal = ${V1 | V2 | \ldots | Vn}$, Sfinal = ${S1 | S2 | \ldots | Sn}$ and Pfinal = ${P1 | P2 | \ldots | Pn}$. If time restriction of the parent clause $tr \neq \epsilon$, we consider this additional time restriction in all the composed automata together with their own time restrictions. Let $\mathcal{C}_1=(N_{\mathcal{C}_1}, n_{0_{\mathcal{C}_1}}, E_{\mathcal{C}_1}, I_{\mathcal{C}_1}), \mathcal{C}_2=(N_{\mathcal{C}_2}, n_{0_{\mathcal{C}_2}}, E_{\mathcal{C}_2}, I_{\mathcal{C}_2}), \ldots,
\mathcal{C}_n=(N_{\mathcal{C}_n}, n_{0_{\mathcal{C}_n}}, E_{\mathcal{C}_n}, I_{\mathcal{C}_n})$. Considering the case where $g \neq \epsilon$ and $tr \neq \epsilon$, and having that $E_{\mathcal{C}_1}*, E_{\mathcal{C}_2}*, \ldots ,E_{\mathcal{C}_n}*$ are the sets of edges considering time restriction $tr$ together with their own time restriction, the resulting automaton is therefore $OR*=(N_{OR*}, n_{0_{OR*}}, E_{OR*}, I_{OR*})$, where:

\begin{compactitem}

\item $N_{OR*}=N_{\mathcal{C}_1} \cup N_{\mathcal{C}_2} \cup \ldots \cup N_{\mathcal{C}_n} \cup \{C_{init},C_{final},C_{skip}\}$.

\item $n_{0_{OR*}}=C_{1init}$.

\item $E_{OR*}=E_{\mathcal{C}_1}* \cup E_{\mathcal{C}_2}* \cup \ldots \cup E_{\mathcal{C}_n}* \cup \{C_{init} \flechalu{}{} C_{1init}, C_{init} \flechalu{}{} C_{2init}, \ldots ,$\\
$C_{init} \flechalu{}{} C_{ninit}\} \cup \{C_{1final} \flechalu{}{} C_{final}, C_{2final} \flechalu{}{} C_{final}, \ldots ,$\\ $C_{nfinal} \flechalu{}{} C_{final}\} \cup \{C_{init} \flechalu{\neg g}{} C_{skip}\}$.

\item $I_{OR*}=I_{\mathcal{C}_1} \cup \{I(n) \equiv x \leq t2 + 1 \, | \, n \in N_{\mathcal{C}_1}-\{C_{1final}\}\} \cup I_{\mathcal{C}_2} \cup \{I(n) \equiv x \leq t2 + 1 \, | \, n \in N_{\mathcal{C}_2}-\{C_{2final}\}\} \cup \ldots \cup I_{\mathcal{C}_n} \cup \{I(n) \equiv x \leq t2 + 1 \, | \, n \in N_{\mathcal{C}_n}-\{C_{nfinal}\}\}$.

\end{compactitem}

This composition of timed automata is shown graphically in Fig.~\ref{automaton_A8-9-10} \textbf{(B)}.

\item If several norms are composed by a \textbf{SEQ-refinement}, that is, we have specified the diagram $(\epsilon,name,g,tr,C_1 \, Seq \, C_2 \, Seq \ldots Seq \, C_n,\epsilon)$, their composition corresponds to an automaton in which the automata corresponding to
each one of the norms are connected in sequence. Let us consider $\mathcal{C}_1, \mathcal{C}_2, \ldots, \mathcal{C}_n$ the automata corresponding to the norms we are composing. The resulting automaton $SEQ*$ preserves the structure of the automata we are composing, adding just one extra node $C_{skip}$ if guard condition of the parent clause $g \neq \epsilon$ and an urgent edge from $C_{1init}$ to this new node guarded with the guard condition negated ($\neg g$). We connect with an urgent edge performing no action the ending node of each automaton in the sequence ($C_{1final}, C_{2final}, \ldots, C_{n-1final}$) with the initial node of the next automaton ($C_{2init}, C_{3init} \ldots, C_{ninit}$). This rule is not applied in the cases of $C_{1init}$ (as there is not previous ending node to connect) and $C_{nfinal}$ (as there is not following initial node to connect). In the initial node of each one of the composed automata we preserve the violation, satisfaction and permission sets of the previous final node. If time restriction of the parent clause $tr \neq \epsilon$, we consider this additional time restriction in all the composed automata together with their own time restrictions. Let $\mathcal{C}_1=(N_{\mathcal{C}_1}, n_{0_{\mathcal{C}_1}}, E_{\mathcal{C}_1}, I_{\mathcal{C}_1}), \mathcal{C}_2=(N_{\mathcal{C}_2}, n_{0_{\mathcal{C}_2}}, E_{\mathcal{C}_2}, I_{\mathcal{C}_2}), \ldots,
\mathcal{C}_n=(N_{\mathcal{C}_n}, n_{0_{\mathcal{C}_n}}, E_{\mathcal{C}_n}, I_{\mathcal{C}_n})$. Considering the case where $g \neq \epsilon$ and $tr \neq \epsilon$, and having that $E_{\mathcal{C}_1}*, E_{\mathcal{C}_2}*, \ldots ,E_{\mathcal{C}_n}*$ are the sets of edges considering time restriction $tr$ together with their own time restriction, the resulting automaton is $SEQ*=(N_{SEQ*}, n_{0_{SEQ*}}, E_{SEQ*}, I_{SEQ*})$, where:

\begin{compactitem}

\item $N_{SEQ*}=N_{\mathcal{C}_1} \cup N_{\mathcal{C}_2} \cup \ldots \cup N_{\mathcal{C}_n} \cup \{C_{skip}\}$.

\item $n_{0_{SEQ*}}=C_{1init}$.

\item $E_{SEQ*}=E_{\mathcal{C}_1}* \cup E_{\mathcal{C}_2}* \cup \ldots \cup E_{\mathcal{C}_n}* \cup \{C_{1init} \flechalu{\neg g}{} C_{skip},C_{1final} \flechalu{}{} C_{2init},$\\
$ C_{2final} \flechalu{}{} C_{3init}, \ldots, C_{n-1final} \flechalu{}{} C_{ninit}\}$.

\item $I_{SEQ*}=I_{\mathcal{C}_1} \cup \{I(n) \equiv x \leq t2 + 1 \, | \, n \in N_{\mathcal{C}_1}-\{C_{1final}\}\} \cup I_{\mathcal{C}_2} \cup \{I(n) \equiv x \leq t2 + 1 \, | \, n \in N_{\mathcal{C}_2}-\{C_{2final}\}\} \cup \ldots \cup I_{\mathcal{C}_n} \cup \{I(n) \equiv x \leq t2 + 1 \, | \, n \in N_{\mathcal{C}_n}-\{C_{nfinal}\}\}$.

\end{compactitem}

This composition of timed automata is shown graphically in Fig.~\ref{automaton_A8-9-10} \textbf{(C)}.

\end{enumerate}

}
\end{definition}

\subsection{Implementation in UPPAAL}

The implementation of the NTAs we have obtained in UPPAAL is quite straightforward as both, the NTA formalism considered by the tool and the NTA formalism that we have considered, are very similar. There are only a few implementation points that need a more detailed explanation:

\begin{compactitem}

\item First, as there is no way in UPPAAL of directly expressing that an edge without synchronisation should be taken without delay, that is, there are no urgent edges, we have to find an alternative way of encoding this behaviour. For this purpose we consider the modelling pattern proposed in \cite{UPPAAL2}. The encoding of urgent edges introduces an extra automaton, that we call $Urgent$, with a single location and a self loop. The self loop synchronises on an urgent channel that we call $urg\_edge$. An edge can now be made urgent by performing the complimentary action.

\item The performance of actions by agents is implemented by means of boolean variables in UPPAAL. We define a boolean variable called $agent\_action$ for each one of the actions considered in the contract. These variables are initialized to \textbf{false} and, when one of the actions is performed by an agent in one of the edges, we update the value of the corresponding variable to \textbf{true}.

\item Finally, the violation, satisfaction and permission sets are implemented in UPPAAL by means of boolean arrays and constant integers with the names of the clauses of the contract containing obligations, prohibitions or permissions. We define an array $V$ for violation, an array $S$ for satisfaction, and an array $P$ for permission, all of them initialized to \textbf{false}. The size of the arrays $V$ and $S$ is equal to the number of obligations and prohibitions in the contract, whereas the size of the array $P$ is equal to the number of permissions. We also define constant integers with the name of the clauses containing obligations and prohibitions, initializing each one of them to a different value (from 0 to the size of the arrays $V$ and $S$ minus 1), and constant integers with the name of the clauses containing permissions, initializing each one of them to a different value (from 0 to the size of the array $P$ minus 1). These constants are used as indexes in the arrays.
When taking a transition where the target node contains at least one modified set (an obligation/prohibition is violated, an obligation/prohibition is satisfied or a permission is made effective), we update to \textbf{true} in the proper array the value of the 
index corresponding to the clause. In the case of repairing an obligation/prohibition violation, the index corresponding to the proper clause in $V$ is set to \textbf{false}.

\end{compactitem}

\subsection{Example: Online Auctioning Process}

\begin{figure}
\vspace{-0.5cm}
\begin{center}
  \includegraphics[width=4.5cm]{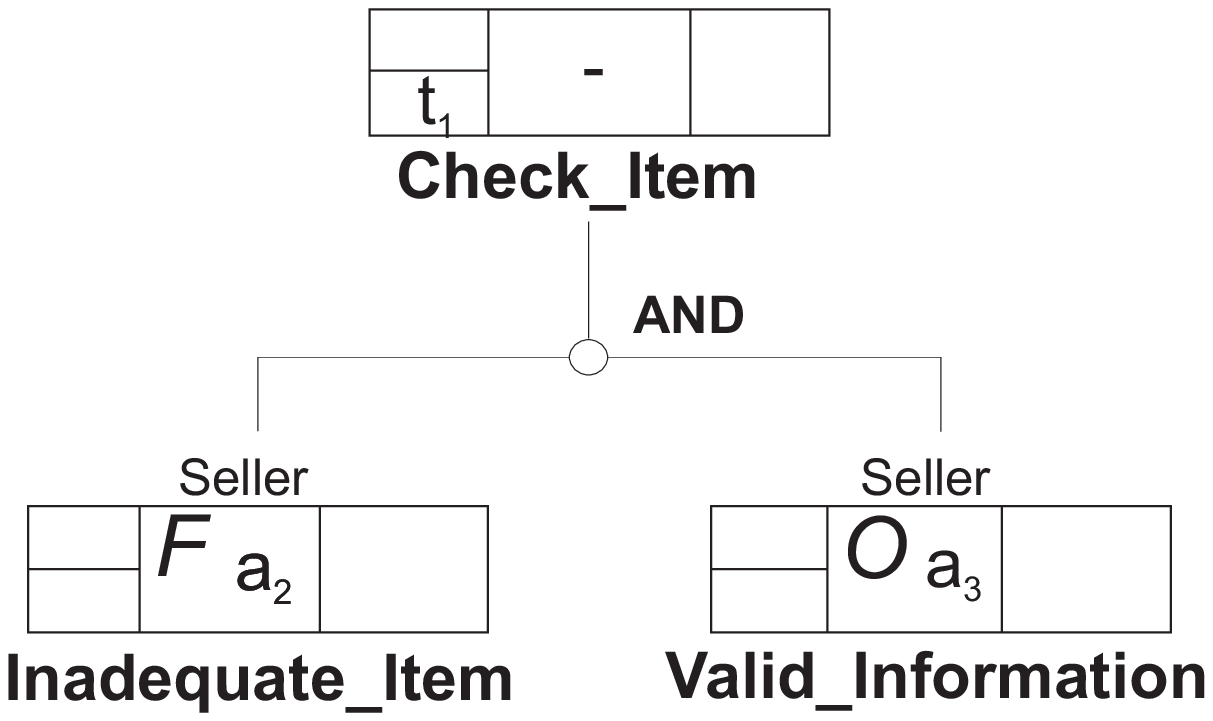}
  \includegraphics[width=5.5cm]{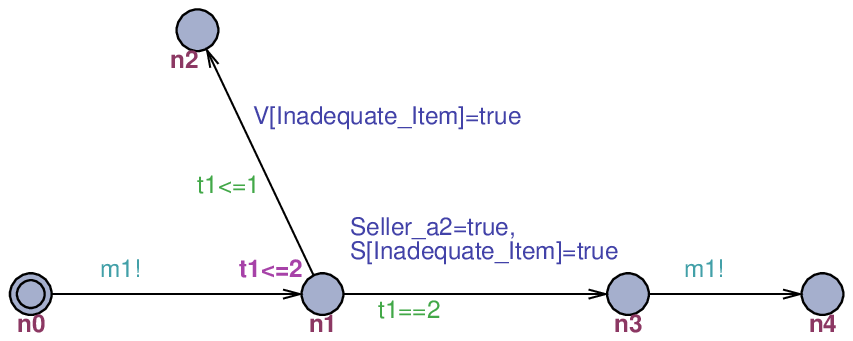}
  \includegraphics[width=5.5cm]{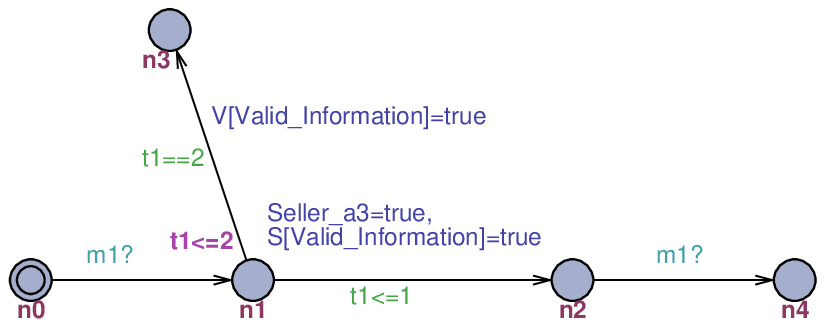}
\end{center}
	\vspace{-0.5cm}
  \caption{Online auctioning process \textit{C-O Diagram} and corresponding NTA in UPPAAL}
  \label{OAP2}
\vspace{-0.5cm}  
\end{figure}

Let us consider part of a contract about an \textit{online auctioning process}. It specifies that at the beginning of the process the \textit{seller} has \textbf{one day} to upload valid information about the item he wants to sell, being forbidden the sale of inadequate items such as replicas of designers items or wild animals. We can identify in this specification an obligation, a prohibition and a real-time constraint affecting both norms. In the representation of this contract as a \textit{C-O Diagram}, that can be seen in the left-hand side of Fig.~\ref{OAP2}, we have a main clause \textit{Check\_Item} including the time restriction \textbf{one day}, denoted as $t_1$. This main clause is decomposed by means of an \textit{AND-refinement}, having on the one hand the clause with the  prohibition, called \textit{Inadequate\_Item} and denoting the action as $a2$, and on the other hand the clause with the obligation, called \textit{Valid\_Information} and denoting the action as $a3$.

By following the \codiag\ semantics, we can obtain an NTA corresponding to the contract. Its implementation in the UPPAAL tool can be seen in the right-hand side of Fig.~\ref{OAP2}, having two automata running in parallel, one corresponding to the prohibition and the other one corresponding to the obligation. Now we can take advantage of all the mechanisms for simulation and formal verification provided by the tool to model-check the contract specification. As this is just a small part of a contract, the properties we can verify here are quite obvious. However, this verification process can be very useful over big contracts, verifying properties such as the violation of clauses when a time constraint expires, the possibility of satisfying the contract without violating any clause, etc. 

For example, in the current NTA we can check the property that if the \textit{seller} takes more that one day ($t1>1$) to upload valid information about the item, the clause \textit{Valid\_Information} is always violated. This property is written as follows in the UPPAAL verifier:

\begin{center}
$A1.n1 \quad and \quad t1>1 --> V[Valid\_Information]==true$
\end{center}

And we obtain that this property is \textbf{satisfied}.

\section{Conclusions}
\label{Conclusions}

In this work we have developed a formal semantics for \codiag\ based on timed automata extended with an ordering of states and edges in order to represent the different deontic modalities. We have also seen how these automata can be implemented in UPPAAL in order to model-check the contract specification, and a small example has been provided.

As future work, we are working on several case studies in order to proove the usefulness of our approach to model-check the specification of complex contracts with real-time constraints. With these case studies we also want to check the complexity of the contracts we can deal with. Finally, we are working on the improvement of the satisfaction rules defined in \cite{Martinez2010} and their relationship with the \codiag\ formal semantics.

\nocite{*}
\bibliographystyle{eptcs}
\bibliography{bibtex}
\end{document}